\documentclass[twocolumn,amsmath,amssymb,aip,pop]{revtex4}

\usepackage{graphicx}
\usepackage{bm}

\usepackage{amsfonts}

\begin{document}
\title{Short-wavelength soliton in a fully degenerate quantum plasma}
\author{Volodymyr M. Lashkin$^{1,2}$}
\email{vlashkin62@gmail.com} \affiliation{$^1$Institute for
Nuclear Research, Pr. Nauki 47, Kyiv 03028, Ukraine}
\affiliation{$^2$Space Research Institute, Pr. Glushkova 40 k.4/1,
Kyiv 03680,  Ukraine}


\begin{abstract}
We present a novel one-dimensional nonlinear evolution equation
governing the dynamics short-wavelength longitudinal waves in a
nonrelativistic fully degenerate quantum plasma using kinetic
equation for the Wigner function. The linear dispersion of the
equation has a form of "zero sound" $\sim k\exp (-k^{2})$, where
$k$ is the wave number, and it strongly differs from previously
known nonlinear evolution equations. We numerically find the
corresponding soliton solutions and demonstrate that the
collisions between three solitons turn out to be elastic resulting
only in  phase shifts.
\end{abstract}

\maketitle

\section{Introduction}

Nonlinear evolution equations are widely used as models to
describe many phenomena in various field of nonlinear science. The
classical examples of these equations are well-known universal
models in dispersive nonlinear media, such as the Korteweg-de
Vries (KdV) and nonlinear Schrodinger (NLS) equations etc.
\cite{Dodd,Zakharov}. A common feature of nonlinear evolution
equations is the presence of dispersion and nonlinearity which in
some cases can effectively balance each other and lead to soliton
formation.

In classical plasma, the most known examples of solitons are the
ion-acoustic soliton described by the KdV equation which
corresponds to the linear dispersion $\omega\sim k^{3}$,  the
Langmuir soliton of the NLS equation (in the subsonic regime) and
Alfv\'{e}n soliton of the derivative NLS equation with the
dispersion $\omega\sim k^{2}$. Here, $\omega$ and $k$ are the
frequency and wave number respectively in suitable dimensionless
variables. At present, a comparatively large number of nonlinear
equations, including multidimensional ones, and, accordingly,
their soliton solutions in classical plasma are known
\cite{Petviashvili1992,Horton_Ichikawa1996,Scorich2010}. Note here
that, generally speaking, most nonlinear equations admitting
soliton solutions are not completely integrable (unlike, for
instance, the KdV, NLS and derivative NLS equations and some
others \cite{Zakharov}), and, therefore, have no the exact
solutions describing the elastic collisions between solitons.

Quantum effects in plasmas are important in the limit of
low-plasma temperature and high-particle number density
\cite{Melrose2008,HaasBook2011}. Such plasmas are ubiquitous in
microelectronic devices, in dense astrophysical plasma, in
microplasmas, and in laser plasmas (for example, see the reviews
\cite{Shukla2011,Vladimirov2011} and references therein). There
are two well-known models for describing the quantum effects in a
plasma. The Wigner and Hartree models are based upon the
Wigner-Poisson and Schr\"{o}dinger-Poisson systems which,
respectively, correspond to the statistical and hydrodynamic
description of the plasma particles. Kinetic models of quantum
plasmas are based on the time evolution equation for one-particle
Wigner function in the mean-field approximation, in which the
self-consistent electrostatic or electromagnetic fields are
described by either Poisson's equation or Maxwell's equations,
respectively. The quantum hydrodynamic (QHD) model generalizes the
fluid model  by including of quantum statistical pressure and
quantum diffraction (the Bohm potential) terms. Later on, the
quantum hydrodynamics model for plasmas was extended to include
magnetic fields \cite{Haas2005} and spin dynamics
\cite{Marklund_Brodin2007}. In the framework of kinetic
description, explicit nonlinear one-dimensional solutions for the
stationary Wigner and Wigner-Poisson equations were presentad in
Ref. \cite{Haas_Wigner}. Based on the kinetic Wigner-Poisson model
of quantum plasma, the kinetic quantum Zakharov equations that
describe nonlinear coupling of Langmuir waves to low frequency
plasma density variations for cases of non-degenerate and
degenerate plasma electrons were obtained in Refs.
\cite{Vladimirov2013,Vladimirov2015}. On the other hand, it is the
QHD model that is most often used to study nonlinear phenomena, in
particular, nonlinear waves and solitons in quantum plasmas
\cite{Shukla2010,Shukla2011}. In this case, the reductive
perturbation technique is usually used to obtain evolution
equations that take into account weak dispersion and weak
nonlinearity. In the framework of the QHD model, the KdV equation
and acoustic solitons in nonrelativistic, unmagnetized cold
quantum plasmas were obtained in Ref. \cite{Haas2003}. In the same
model, nonlinear periodic waves (cnoidal waves), which generalize
soliton solutions of the KdV equation to the case of periodic
boundary conditions, were investigated in Ref. \cite{Haas2014}. In
Ref. \cite{Haas2015} linear ion-acoustic waves and ion-acoustic
solitons of the KdV equation were studied in the QHD model for
nonrelativistic, unmagnetized quantum plasma with electrons with
an arbitrary degeneracy degree. Under this, the equation of state
for electrons follows from a local Fermi-Dirac distribution
function and applies equally well both to fully degenerate and
classical, nondegenerate limits. In the long-wavelength limit, the
results agree with quantum kinetic theory. The KdV equation and
the corresponding soliton solutions were obtained also in a
magnetized quantum plasma in the form of the magnetoacoustic
solitons \cite{Hussain2011}, and in a plasma with relativistically
degenerate electrons \cite{Masood2011}. Solitons of the
Kadomtsev-Petviashvili equation, which is a multidimensional
generalization of the KdV equation for an unmagnetized or weakly
magnetized plasma \cite{Petviashvili1992}, were obtained for
quantum plasmas in Ref. \cite{Sahu2013}. The same equation was
derived for describing ion-acoustic nonlinear waves in cold
quantum electron-positron-ion plasma \cite{Mustaq2007} and for
magnetoacoustic solitons in magnetized quantum plasma
\cite{Wang2016}. The modified KdV equation (the KdV with qubic
nonlinearity, and also completely integrable) in a quantum plasma
and its soliton were also obtained in Ref. \cite{Hosen2016}.
Solitons of the Zakharov-Kuznetsov equation
\cite{Petviashvili1992,Kuznetsov1974}, which is a multidimensional
generalization of the KdV equation for magnetized plasma, were
considered for cold quantum plasma in Refs.
\cite{Moslem2007,Sabry2008,Khan2008,Yan2009} in the form of the
usual ion-acoustic solitons, explosive solitons and nonlinear
periodic waves in terms of the Jacobi elliptic functions and in
Ref. \cite{Haas2016} for magnetized quantum plasma with arbitrary
degeneracy of electrons. Recently, a nonlinear string equation
(one of the forms of the Boussinesq equation) with the linear
dispersion $\omega^{2}\sim -k^{2}+k^{4}$ and quadratic
nonlinearity for describing quantum two-stream instability  has
been suggested in Ref. \cite{Lashkin2020}. The corresponding
analytical solutions have the form of the blow-up solitons .

As is known, the considered hydrodynamic models are valid only in
the long-wavelength case $k\ll 1$ (the wave number is normalized
to the Fermi-Debye length) \cite{HaasBook2011}. Then, the
appearance of solitons is due to the balance of weak dispersion
$k^{3}\ll 1$ for the KdV equation, and $k^{2}\ll 1$ for the NLS,
for example, and weak nonlinearity (quadratic and cubic
respectively). The goal of this paper is to derive a novel
evolution equation describing short-wavelength ($k\gg 1$)
nonlinear waves in a nonrelativistic fully degenerate  quantum
plasma using the kinetic approach. Despite the specific nature of
the dispersion in the short-wavelength limit (the dispersion of
"zero sound") $\omega\sim k\exp (-k^{2})$, which has no
counterpart in classic plasmas, we show that balance between the
weak dispersion and weak quadratic nonlinearity lead to the
formation of solitons. Moreover, we show that collisions between
even three solitons are elastic.

The paper is organized as follows. In Sec. II, we present our
model nonlinear equation that governs the dynamics of longitudinal
waves in a fully degenerate quantum plasma in the short-wavelength
limit. The soliton solutions and elastic collisions between
solitons are presented in Sec. III. Finally, Sec. IV concludes the
paper.

\section{Model equation}

Dielectric functions and dispersion relations of quantum
degenerate plasmas have been calculated using kinetic theory
\cite{Klimontovich1952,Klimontovich1960,Melrose2008}. In the
semiclassical limit $\hbar k\ll p_{F}=\hbar
(3\pi^{2}n_{0})^{1/3}$, the longitudinal dielectric function in a
plasma with completely degenerate electrons in the
zero-temperature limit reads
\begin{gather}
\varepsilon
(\omega,\mathbf{k})=1+\frac{3\omega_{p}^{2}}{k^{2}v_{F}^{2}}\left[1-\frac{\omega}{2kv_{F}}
\ln\left(\frac{\omega+kv_{F}}{\omega-kv_{F}}\right)\right]
\nonumber \\
+i\pi\frac{\omega}{2kv_{F}}\theta (k^{2}v_{F}^{2}-\omega^{2}),
\label{permittivity}
\end{gather}
where $\theta(x)$ is the Heaviside step function,  $\omega$ and
$\mathbf{k}$ are the frequency and the wave vector respectively
with $k=|\mathbf{k}|$, $\omega_{p}=\sqrt{4\pi e^{2}n_{0}/m}$ is
the electron plasma frequency, $v_{F}=\hbar
(3\pi^{2}n_{0})^{1/3}/m$ is the electron Fermi speed, $n_{0}$ is
the equilibrium plasma density, $m$ and $-e$ are the electron mass
and charge, respectively, $\hbar$ is the Planck constant divided
by $2\pi$. Analytical expressions for the wave dispersion can be
obtained from the dispersion equation $\varepsilon
(\omega,\mathbf{k})=0$ in the two limiting cases
\cite{Klimontovich1952,Klimontovich1960,
Melrose2008,Landau_Lifshit10}. In the long-wavelength limit
$\omega_{p}\gg kv_{F}$, the dispersion relation for longitudinal
waves is
\begin{equation}
\label{displong}
\omega_{\mathbf{k}}=\omega_{p}\left(1+\frac{3}{10}k^{2}v_{F}^{2}\right),
\end{equation}
and similar to the dispersion of Langmuir waves in classical
electron-ion plasmas. This dispersion has the optical type (i. e.
$\omega_{\mathbf{k}}\rightarrow \omega_{c}$ as $k\rightarrow 0$,
where $\omega_{c}$ is the cut-off frequency) and for the weak
nonlinearity is usually associated with the four-wave interaction
and NLS-like equations. In particular, in the one-dimensional case
balance between dispersion in Eq. (\ref{displong}) and cubic
nonlinearity leads to a soliton similar to the well-known Langmuir
soliton. In the opposite case, i. e. in the short-wavelength limit
$\omega_{p}\ll kv_{F}$, the dispersion relation for the
longitudinal wave has the form
\begin{equation}
\label{dispersion}
\omega_{\mathbf{k}}=kv_{F}\left[1+2\exp\left(-\frac{2}{3}\frac{k^{2}v_{F}^{2}}
{\omega_{p}^{2}}-2\right)\right].
\end{equation}
There is no counterpart of the dispersion relation Eq.
(\ref{dispersion}) in classical plasmas. Note that the dispersion
is of the acoustic type (i.e. $\omega_{\mathbf{k}}\rightarrow 0$
as $k\rightarrow 0$). The Landau damping is absent in both cases
Eqs. (\ref{displong}) and (\ref{dispersion}) since
$\omega/k>v_{F}$. We address a new type of nonlinear evolution
equation which is characterized by the linear dispersion Eq.
(\ref{dispersion}).

Obtaining of a nonlinear equation with the dispersion Eq.
(\ref{dispersion}) requires an essentially kinetic description. In
kinetic theory, the response of a plasma to longitudinal (i.e.
electrostatic) wave fields is described by the linear response and
a hierarchy of nonlinear susceptibilities
\cite{Sitenko,Tsytovich1995}. In a quantum plasma, this approach
was used to derive the Zakharov equations taking into account
quantum corrections \cite{Vladimirov2013}. In the following we
consider the one-dimensional case and use the notation
\begin{equation}
\sum_{q=q_{1}+q_{2}} \cdots \rightarrow\int \cdots \,\delta
(q-q_{1}-q_{2})\frac{dq_{1}}{(2\pi)^{2}}\frac{dq_{2}}{(2\pi)^{2}},
\end{equation}
where $\delta (x)$ is the Dirac delta function and $q=(\omega,k)$.
In the one-dimensional space the Wigner kinetic equation
\cite{Moyal1949,Klimontovich1952,Tatarskii1983} for the quantum
electron distribution function (Wigner function) $F(x,v,t)$ can be
written as
\begin{gather}
\frac{\partial F}{\partial t}+v\frac{\partial F}{\partial
x}=-\frac{iem}{2\pi\hbar^{2}}\int\int d\lambda dv^{'}\exp
\left[i\frac{m}{\hbar}(v-v^{'})\lambda\right]  \nonumber \\
\times\left[\varphi\left(x+\frac{\lambda}{2},t\right)
-\varphi\left(x-\frac{\lambda}{2},t\right)\right]F(x,v^{'},t),
\label{basic_tatar}
\end{gather}
where $\varphi$ is the electrostatic potential. In the momentum
space eq. (\ref{basic_tatar}) can be written as
\begin{gather}
(\omega-kv) f_{q}(v)=\frac{em}{2\pi\hbar^{2}}\int\int d\lambda
dv^{'}\exp
\left[i\frac{m}{\hbar}(v-v^{'})\lambda\right]  \nonumber \\
\times\left[\left(\mathrm{e}^{ik\lambda/2}-\mathrm{e}^{-ik\lambda/2}\right)\varphi_{q}f^{(0)}(v^{'})
 \right. \nonumber \\ \left. +\sum_{q=q_{1}+q_{2}}\left(\mathrm{e}^{ik_{1}\lambda/2}-\mathrm{e}^{-ik_{1}\lambda/2}\right)
\varphi_{q_{1}}f_{q_{2}}(v^{'})\right] \label{basic_kinetic1}
\end{gather}
where $f_{q}(v)$ is the deviation of the electron distribution
function from the equilibrium one $f^{(0)}(v)$, and $\varphi$ is
the electrostatic potential. The distribution function
$f^{(0)}(v)$ is normalized to the equilibrium plasma density ,
$\int f^{(0)}(v)dv=n_{0}$. Integrating over $\lambda$ in eq.
(\ref{basic_kinetic1}) yields
\begin{equation}
\exp \left[i\frac{m}{\hbar}(v-v^{'})\pm\frac{ik}{2}\right]\lambda
\rightarrow 2\pi\delta
\left[\frac{m}{\hbar}(v-v^{'})\pm\frac{k}{2}\right],
\end{equation}
and then integrating over $v^{'}$ yields
\begin{gather}
(\omega-kv)
f_{q}(v)=\frac{e}{\hbar}\varphi_{q}\left[f^{(0)}\left(v+\frac{\hbar
k}{2m}\right)-f^{(0)}\left(v-\frac{\hbar k}{2m}\right)\right]
\nonumber \\
+\frac{e}{\hbar}\sum_{q=q_{1}+q_{2}}\varphi_{q_{1}}\left[f_{q_{2}}\left(v+\frac{\hbar
k_{1}}{2m}\right)-f_{q_{2}}\left(v-\frac{\hbar
k_{1}}{2m}\right)\right].
 \label{basic_kinetic}
\end{gather}
We present the function $f_{q}(v)$ as a series in powers of the
field strength (i. e. $f_{q}^{(n)}\sim \varphi^{n}$)
\begin{equation}
\label{series} f_{q}(v)=\sum_{n=1}^{\infty}f_{q}^{(n)}(v).
\end{equation}
In the linear approximation from Eqs. (\ref{basic_kinetic}) and
(\ref{series}) we have
\begin{equation}
f_{q}^{(1)}=\frac{e\varphi_{q}}{\hbar
(\omega-kv)}\left[f^{(0)}\left(v+\frac{\hbar
k}{2m}\right)-f^{(0)}\left(v-\frac{\hbar k}{2m}\right)\right],
\end{equation}
and then one can write the recurrence relation
\begin{gather}
f_{q}^{(n)}=\frac{e}{\hbar
(\omega-kv)}\sum_{q=q_{1}+q_{2}}\varphi_{q_{1}}\left[f_{q_{2}}^{(n-1)}\left(v+\frac{\hbar
k_{1}}{2m}\right) \right. \nonumber \\
 \left. -f_{q_{2}}^{(n-1)}\left(v-\frac{\hbar
k_{1}}{2m}\right)\right]. \label{recur1}
\end{gather}
For the nonlinear terms ($n \geqslant 2$) we use the
quasiclassical approximation and in the limit $\hbar k/(2m)\ll v$
in Eq. (\ref{recur1})  one can expand
\begin{equation}
f_{q_{2}}^{(n-1)}\left(v\pm\frac{\hbar k_{1}}{2m}\right)\approx
f_{q_{2}}^{(n-1)}(v)\pm\frac{\partial f_{q_{2}}^{(n-1)}}{\partial
v}\frac{\hbar k_{1}}{2m}
\end{equation}
whence we get
\begin{equation}
\label{recur2} f_{q}^{(n)}=\frac{e}{m(\omega-kv)}
\sum_{q=q_{1}+q_{2}}k_{1}\varphi_{q_{1}}\frac{\partial
f_{q_{2}}^{(n-1)}}{\partial v}.
\end{equation}
Retaining terms in Eq. (\ref{series}) up to second order in the
wave fields and substituting $f_{q}$ into the Poisson equation
\begin{equation}
k^{2}\varphi_{q}=-4\pi e \int f_{q}(v)dv ,
\end{equation}
where  the ion contribution is omitted, we get
\begin{equation}
\label{nonlin_eq1} \varepsilon_{q}\varphi_{q}=\sum_{q=q_{1}+q_{2}}
V_{q_{1},q_{2}}\varphi _{q_{1}}\varphi_{q_{2}},
\end{equation}
where
\begin{equation}
\label{linear_responce} \varepsilon_{q}=1+\frac{4\pi e^{2}}{\hbar
k^{2}}\int\frac{\left[f^{(0)}\left(v+\frac{\hbar
k}{2m}\right)-f^{(0)}\left(v-\frac{\hbar
k}{2m}\right)\right]}{(\omega-kv)}dv,
\end{equation}
or, after suitable change of variables \cite{Shukla2010}
\begin{equation}
\label{linear_responce1} \varepsilon_{q}=1-\frac{4\pi
e^{2}}{m}\int\frac{f^{(0)}}{(\omega-kv)^{2}-\hbar^{2}k^{4}/(4m^{2})}dv,
\end{equation}
is the linear electron dielectric response function from which one
can obtain Eq. (\ref{permittivity}) and, in particular, the linear
dispersion law Eq. (\ref{dispersion}), and the interaction matrix
element (the nonlinear dielectric susceptibility)
$V_{q_{1},q_{2}}$ is determined by
\begin{gather}
V_{q_{1},q_{2}}=-\frac{e}{2m}
\frac{\omega^{2}_{p}}{n_{0}k^{2}}\int
\frac{k_{1}}{[(\omega_{1}+\omega_{2})-(k_{1}+k_{2})
v]} \nonumber \\
\times\frac{\partial}{\partial
v}\frac{k_{2}}{(\omega_{2}-k_{2}v)}\frac{\partial
f^{(0)}}{\partial v}dv +(\omega_{1},k_{1} \rightleftarrows
\omega_{2},k_{2}). \label{matrix1}
\end{gather}
Note that the expression (\ref{matrix1}) for the interaction
matrix element $V_{q_{1},q_{2}}$ is written in a symmetrized form.
Singularities in the denominators in Eqs. (\ref{linear_responce})
and (\ref{matrix1})  are avoided, as usual, using Landau's rule by
replacing  $\omega\rightarrow\omega+i0$ and then
\begin{equation}
(\omega-kv)^{-1}=\mathcal{P}(\omega-kv)^{-1}-i\pi\delta
(\omega-kv), \label{Landau_lin}
\end{equation}
\begin{gather}
[\omega_{1}+\omega_{2}-(k_{1}+k_{2})v]^{-1}=\mathcal{P}[\omega_{1}+\omega_{2}-(k_{1}+k_{2})v]^{-1}
\nonumber \\ -i\pi\delta [\omega_{1}+\omega_{2}-(k_{1}+k_{2})v],
\label{Landau_nonlin}
\end{gather}
where $\mathcal{P}$ is the principal value of the integrals.
Imaginary parts in Eqs. (\ref{Landau_lin}) and
(\ref{Landau_nonlin}) account for the linear and nonlinear Landau
damping respectively. As noted above, the linear Landau damping in
a fully degenerate quantum plasma for the wave with dispersion Eq.
(\ref{dispersion}) is absent, but the nonlinear Landau damping due
to the interaction of the beat waves with plasma particles is
possible \cite{Brodin2015}. In this work we neglect the nonlinear
Landau damping and only the principal value of the integrals is
understood, although the corresponding damping term can be easily
obtained from Eq. (\ref{Landau_nonlin}) in the same way as the
nonlinear Landau damping is obtained in the kinetic derivation of
the NLS equation for Langmuir waves in classic plasma
\cite{Horton_Ichikawa1996}. After two partial integrations in Eq.
(\ref{matrix1}), one can write
\begin{gather}
V_{q_{1},q_{2}}=-\frac{e}{2m}
\frac{\omega^{2}_{p}}{n_{0}k^{2}}\int \left[
\frac{2k^{2}k_{1}k_{2}}{(\omega-kv)^{3} (\omega_{2}-k_{2}v)}
\right.
 \nonumber \\
 \left. +\frac{kk_{1}k_{2}^{2}}{(\omega-
 kv)^{2}(\omega_{2}-k_{2}v)^{2}}\right]f^{(0)}
dv+(\omega_{1},k_{1}\rightleftarrows\omega_{2},k_{2}),
\label{matrix2}
\end{gather}
where $\omega=\omega_{1}+\omega_{2}$ and $k=k_{1}+k_{2}$.
Expanding $\varepsilon(\omega,k)$ given by Eq.
(\ref{permittivity}) near the $\omega_{k}$ determined by Eq.
(\ref{dispersion}) yields
\begin{equation}
\label{expansion} \varepsilon(\omega,k)=\varepsilon(\omega_{k},k)+
\varepsilon^{'}(\omega_{k})(\omega-\omega_{k})
\end{equation}
where $\varepsilon^{'}(\omega_{k})\equiv\partial \varepsilon
(\omega)/\partial\omega\mid_{\omega=\omega_{k}}$ and in the
leading order one can get
\begin{equation}
\label{epsilon_der}
\varepsilon^{'}(\omega_{k})=\frac{3\omega_{p}^{2}}{4k^{3}v_{F}^{3}}
\exp\left(\frac{2}{3}\frac{k^{2}v_{F}^{2}}{\omega^{2}_{p}}+2\right).
\end{equation}
After substituting Eq. (\ref{expansion}) into Eq.
(\ref{nonlin_eq1})  we have
\begin{equation}
\label{nonlin_eq_basic}
(\omega-\omega_{k})\varphi_{q}=\frac{1}{\varepsilon^{'}(\omega_{k})}
\sum_{q=q_{1}+q_{2}} V_{q_{1},q_{2}}\varphi
_{q_{1}}\varphi_{q_{2}}.
\end{equation}
Note that the direct substitution of the equilibrium distribution
function for the one-dimensional fully degenerate electrons
$\partial f^{(0)}/\partial v=-n_{0}/(2v_{F})\delta (v_{F}-|v|)$
into Eq. (\ref{matrix1}) after one partial integration is in
excess of accuracy, since it takes into account dispersion terms
in the nonlinearity which correspond to $k_{1,2}v$ in the
denominators. Thus, we neglect these terms in Eq. (\ref{matrix2})
and in the leading order the nonlinear dielectric susceptibility
Eq. (\ref{matrix2}) becomes
\begin{equation}
\label{matrix3}
V_{_{q_{1}},q_{2}}=-\frac{e}{2m}\frac{\omega_{p}^{2}}{k^{2}}\left\{\frac{2k^{2}k_{1}k_{2}}
{\omega^{3}\omega_{2}}
+\frac{kk_{1}k_{2}^{2}}{\omega^{2}\omega_{2}^{2}}
+(\omega_{1},k_{1}\rightleftarrows\omega_{2},k_{2})\right\}.
\end{equation}
Essentially, that the wave dispersion in Eq. (\ref{dispersion})
has an acoustic type and in the leading term satisfies the
three-wave resonance condition
\begin{equation}
\label{resonance} \omega_{k}=\omega_{k_{1}}+\omega_{k_{2}}, \,
k=k_{1}+k_{2}.
\end{equation}
In particular, this means that this condition, together with
taking into account only the quadratic nonlinearity, ensures the
validity of the successive approximation in Eq. (\ref{series}),
and this is equivalent \cite{Sitenko,Sitenko1973} to the
multi-time-scale perturbation expansion, i. e. the secular terms
are removed automatically. Taking into account Eqs.
(\ref{dispersion}) and (\ref{resonance}) when calculating Eq.
(\ref{matrix3}), then substituting Eqs. (\ref{epsilon_der}) and
(\ref{matrix3}) into Eq. (\ref{nonlin_eq_basic}), and introducing
the slow time scale $\Omega=\omega-kv_{F}$ which balances the
dispersion in Eq. (\ref{dispersion}) (compare, for example,
kinetic derivation of the KdV equation in Ref.
\cite{Scorich2010}), we finally get
\begin{gather}
\left[\Omega -
2kv_{F}\exp\left(-\frac{2}{3}\frac{k^{2}v_{F}^{2}}{\omega^{2}_{p}}-2\right)\right]\varphi_{q}
\nonumber \\
=-\frac{2e}{mv_{F}}k
\exp\left(-\frac{2}{3}\frac{k^{2}v_{F}^{2}}{\omega^{2}_{p}}-2\right)
\sum_{q=q_{1}+q_{2}}\varphi_{q_{1}}\varphi_{q_{2}} \label{main3}
\end{gather}
After rescaling
\begin{equation}
k\rightarrow \frac{v_{F}}{\omega_{p}}k, \, \Omega\rightarrow
\frac{\exp (2)}{2\omega_{p}}\Omega, \, \Phi\rightarrow
-\frac{4e}{mv_{F}^{2}}\varphi,
\end{equation}
equation (\ref{main3}) can be written in the dimensionless form
\begin{equation}
\label{main33} \left[\Omega -
k\exp\left(-\frac{2}{3}k^{2}\right)\right]\Phi_{q} =k
\exp\left(-\frac{2}{3}k^{2}\right)
\sum_{q=q_{1}+q_{2}}\Phi_{q_{1}}\Phi_{q_{2}}.
\end{equation}
By introducing the operator $\hat{L}$ acting in the physical space
as
\begin{equation}
\hat{L}f(x)=\int ik\exp (-2k^{2}/3)\mathrm{e}^{-ikx}
\hat{f}(k)\,dk,
\end{equation}
where $f(x)$ is an arbitrary function and $\hat{f}(k)$ is its
Fourier transform, and using the convolution identity
\begin{equation}
\label{convol} (fg)_{k}=\int\hat{f}_{k_{1}}\hat{g}_{k_{2}}\delta
(k-k_{1}-k_{2})\,dk_{1}dk_{2}.
\end{equation}
one can write Eq. (\ref{main33}) in the physical space as
\begin{equation}
\label{main4}
\partial_{t}\Phi +\hat{L}\Phi +\hat{L}\Phi^{2}=0,
\end{equation}
so that the nonlinearity has a nonlocal character. It is seen that
"motionless" ($\partial_{t}=0$) solutions is not possible. Note
also that in the considered short-wavelength case $k>1$,  Eq.
(\ref{main33}) can not be simplified by any expansion in $k$.

\section{Soliton solution and collisions between solitons}

We look for stationary traveling solutions of Eq. (\ref{main4}) of
the form $\Phi(x,t)=\Phi(x-vt)$, where $v$ is the velocity of
propagation in the $x$ direction. In the Fourier space the
stationary solution corresponds to $\Phi_{q}=\Phi_{k}\delta
(\Omega-kv)$ and from Eq. (\ref{main33}) we have
\begin{equation}
\label{main5} \left[v-\exp
\left(-2k^{2}/3\right)\right]\Phi_{k}=\exp
\left(-2k^{2}/3\right)\sum_{k=k_{1}+k_{2}}\Phi_{k_{1}}\Phi_{k_{2}}.
\end{equation}
\begin{figure}
\includegraphics[width=3.2in]{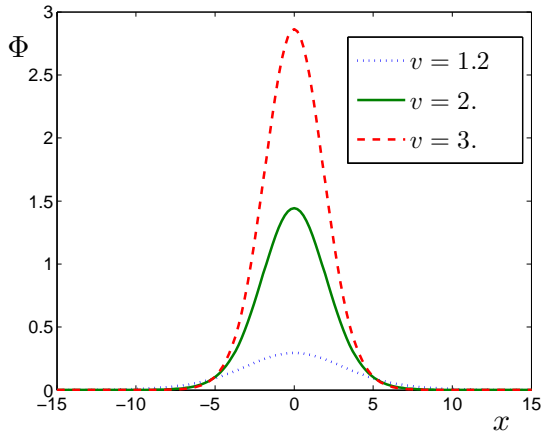}
\caption{\label{fig1}   Examples of the soliton solutions of Eq.
(\ref{main4}) with the different velocities $v$: dotted curve --
$v=1.2$, solid curve -- $v=2$, dashed curve -- $v=3$. }
\end{figure}
Finding an analytical solution of Eq. (\ref{main5}) apparently
does not seem possible but one can find soliton solutions
numerically using the Petviashvili method
\cite{Petviashvili76,Lakoba07}. Equation (\ref{main5}) are written
in the form
\begin{equation}
G_{k}\Phi_{k}=B_{k},
\end{equation}
where $G_{k}=\left[v-\exp \left(-2k^{2}/3\right)\right]$ and
$B_{k}$ accounts for the nonlinear term. Then the Petviashvili
iteration procedure at the $n$-th iteration is
\begin{equation}
\Phi_{k}^{(n+1)}=sG_{k}^{-1}B_{k}^{(n)},
\end{equation}
where $s$ is the so called stabilizing factor determined by
\begin{equation}
s=\left(\frac{\int |\Phi_{k}^{(n)}|^{2}dk}{\int
\Phi_{k}^{\ast,(n)}G_{k}^{-1}B_{k}^{(n)} dk}\right)^{\gamma}.
\end{equation}
and $\gamma=2$ for the quadratic nonlinearity, the parenthetic
superscript denotes the iteration step index. Nonlinear terms at
each step  were calculated by using Eq. (\ref{convol}).
\begin{figure}
\includegraphics[width=3.in]{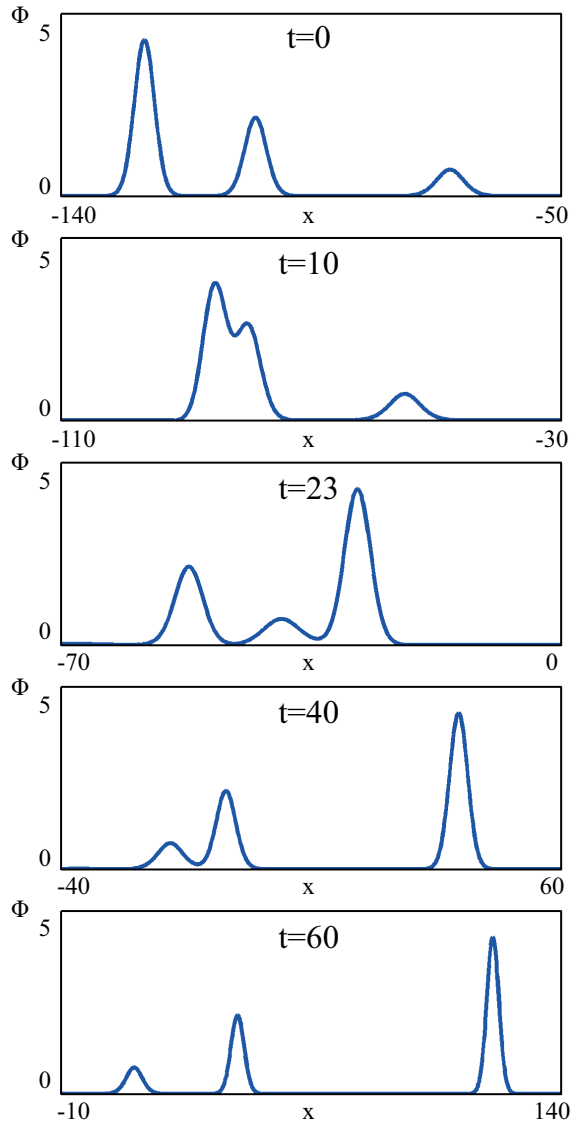}
\caption{\label{fig2}  A typical example of an elastic collision
between three  solitons, with the velocities $v_{1}=4.$,
$v_{2}=2.5$ and $v_{2}=1.5$. The corresponding initial locations
at the moment $t=0$ are $x_{1}=-125.$, $x_{2}=-105.$ and
$x_{2}=-70.$.}
\end{figure}
The procedure always converges to the nonlinear ground state, i.
e. soliton, regardless of the initial guess. Moreover, the rate of
convergence is almost independent of the initial approximation. We
used $\Phi (x)=\exp (-x^{2})$ as the initial guess in all runs.
Iterations rapidly converge to a soliton solution provided that
$v>1$. In physical variables, the soliton velocity should satisfy
the condition $v>2\exp (-2)v_{F}\sim 0.27v_{F}$. Note that for the
group velocity $v_{gr}=\partial\omega/\partial k$ of linear waves
with dispersion Eq. (\ref{dispersion}) (taking into account
$kv_{F}/\omega_{p}\gg 1$) is
\begin{equation}
v_{gr}=v_{F}\left[1-\frac{8k^{2}v_{F}^{2}}{3\exp
(2)\omega_{p}^{2}}\exp\left(-\frac{2}{3}\frac{k^{2}v_{F}^{2}}
{\omega_{p}^{2}}-2\right)\right]
\end{equation}
and $v_{gr}<v_{F}$. Simulations show that the soliton amplitude
grows linearly with increasing the velocity $v$, as for the
solitons of the KdV equation (but not for the Langmuir solitons of
the NLS equation, for which the soliton velocity and amplitude are
independent parameters). Examples of the solitons with different
velocities (amplitudes) are presented in Fig.~\ref{fig1}.

We note that, generally speaking, collisions between solitons in
nonintegrable models can be almost elastic under certain
conditions, for instance, if the soliton amplitudes and velocities
are sufficiently close to each other
\cite{Makhankov1978,Kivshar1989}. To study the time evolution of
the solitons under their collisions, we numerically solve the
nonlinear equation (\ref{main4}) with the initial conditions given
by a superposition of $N\leqslant 3$ soliton solutions
\begin{equation}
\Phi(x,t)=\sum_{i=1}^{N}\Phi_{i}(x-x_{i},t)
\end{equation}
at the time $t=0$, where $\Phi_{i}$ correspond numerically found
(up to machine accuracy) soliton solutions with essentially
different velocities $v_{i}$. The time integration is performed by
a fourth order Runge-Kutta method with the variable time step and
local error control. The periodic boundary conditions are assumed.
The linear and nonlinear terms are computed in spectral space. The
simulations have been performed for various values of soliton
velocities (and, therefore, amplitudes) both for two soliton
($N=2$) and three ($N=3$) soliton collisions. An example of the
elastic collision between three solitons with the velocities
$v_{1}=4.$, $v_{2}=2.5$ and $v_{2}=1.5$ is shown in
Fig.~\ref{fig2}. In particular, it can be seen that at the time
$t=23$ in the inset Fig.~\ref{fig2} all three solitons undergo
distortion simultaneously so that two distant solitons feel each
other through an intermediate soliton -- this is a typical
many-soliton effect. Then, the solitons fully reconstruct their
initial form without any emitting wakes of radiation ($t=60$),
resulting only in phase shifts. The overall picture closely
resembles the elastic soliton collisions in the integrable models
\cite{Zakharov,Ablowitz1987,Tahtadjan1987}.

\section{Conclusions}

In conclusion, we have derived the nonlinear evolution equation
governing dynamics of the short-wavelength longitudinal waves in
the fully degenerate quantum plasma with the "zero-sound"
dispersion and numerically found the soliton solutions. By
numerical simulation we have shown that soliton collisions are
elastic.

The elastic collisions between three solitons might suggest that
equation (\ref{main4}) has exact $N$-soliton solutions and is
completely integrable just like for KdV equation and others
\cite{Zakharov,Ablowitz1987,Tahtadjan1987}, but this is most
likely not the case. In the inverse scattering transform approach
there exists a relationship between some function
$\hat{\omega}(\lambda)$, where $\lambda$ is the spectral
parameter, and the dispersion relation $\omega (k)$ of the
linearized equation \cite{Ablowitz1987}. In all known cases
$\hat{\omega}(\lambda)$ is the rational function of $\lambda$
though the associated spectral problem may involve meromorphic
functions of the spectral parameter $\lambda$ like the elliptic
Jacobi functions, as in the case of the Landau-Lifshitz equation
\cite{Tahtadjan1987}. In any case, the integrability of equation
(\ref{main4}) seems to be an open question.

Note that the dispersion relation Eq. (\ref{dispersion}) is the
same as in classical nondegenerate ultrarelativistic plasmas, i.
e. when the plasma particle temperature significantly exceeds the
particle rest mass
\cite{Landau_Lifshit10,Melrose2008,Silin1960,Thoma09}, in the
short-wavelength limit $k\gg \omega_{p}/c$ with the replacement
$v_{F}\rightarrow c$, where $c$ is the speed of light and
$\omega_{p}\rightarrow \sqrt{4\pi e^{2}c^{2}n_{0}/(3T)}$ is the
ultrarelativistic electron Langmuir frequency and $T$ is the
electron temperature. In reality, ultrarelativistic plasma
consists of electrons and positrons and, in most cases, a small
impurity of nonrelativistic ions. A preliminary consideration
shows that for pure electron-positron plasma in the thermal
equilibrium (electron and positron temperatures are equal),
quadratic nonlinearity vanishes identically. The situation changes
drastically if ions are present, then one can obtain an equation
similar to Eq. (\ref{main4}). Under this, the dominant nonlinear
term comes from the electrons and positrons while the ion
contribution in nonlinearity is negligible. The detailed analysis
will be presented elsewhere.

\section{Data availability}
Data sharing is not applicable to this article as no new data were
created or analyzed in this study.

\end{document}